\documentclass[10pt]{IEEEtran}

\usepackage[utf8]{inputenc}
\usepackage{graphicx}
\usepackage{url}
\usepackage{color}
\usepackage{hyperref}
\usepackage{listings}
\usepackage[dvipsnames]{xcolor}
\usepackage{tikz} 
\usetikzlibrary{shapes,calc,shadows,arrows}
\usepackage{subcaption}
\usepackage{verbatim}
\usepackage{enumitem}

\begin{document}

\title{Logging Practices with Mobile Analytics:\\An Empirical Study on Firebase}


\author{\IEEEauthorblockN{Julian Harty\IEEEauthorrefmark{1}\textsuperscript{1}, Haonan Zhang\IEEEauthorrefmark{1}\textsuperscript{2}, Lili Wei\textsuperscript{3}, Luca Pascarella\textsuperscript{4}, Maurício Aniche\textsuperscript{5}, Weiyi Shang\textsuperscript{2}}\\\IEEEauthorblockA{\textsuperscript{1}The Open University, Commercetest Ltd, \textsuperscript{2}Concordia University,\\\textsuperscript{3}The Hong Kong University of Science and Technology,\\\textsuperscript{4}Università della Svizzera Italiana, \textsuperscript{5}Delft University of Technology}}

\maketitle
\begingroup\renewcommand\thefootnote{\IEEEauthorrefmark{1}}
\footnotetext{The first two authors contributed equally to this work.}
\endgroup
\thispagestyle{plain}
\pagestyle{plain}
\begin{abstract}

Software logs are of great value in both industrial and open-source projects. Mobile analytics logging enables developers to collect logs remotely from their apps running on end user devices at the cost of recording and transmitting logs across the Internet to a centralised infrastructure. 

This paper makes a first step in characterising logging practices  with a widely adopted mobile analytics logging library, namely Firebase Analytics. 
We provide an empirical evaluation of the use of Firebase Analytics in 57 open-source Android applications by studying the evolution of code-bases to understand: a) the needs-in-common that push practitioners to adopt logging practices on mobile devices, and b) the differences in the ways developers use local and remote logging.

Our results indicate mobile analytics logs are less pervasive and less maintained than traditional logging code. 
Based on our analysis, we believe logging using mobile analytics is more user centered compared to traditional logging, where the latter is mainly used to record information for debugging purposes. 

\end{abstract}

\section{Introduction}

Logs are valuable and sometimes the only available source of runtime information of software systems~\cite{Yuan:2010:SED:1736020.1736038}. Produced by logging statements, logs play important roles in the daily tasks of developers and other software practitioners~\cite{Barik2016, hengtse}. The rich information in logs has been widely leveraged in both practice and research to accomplish challenging tasks in both software development and operation, including system comprehension~\cite{fu2013contextual, Shang:2013:ADB:2486788.2486842,wieman2017experience}, anomaly detection~\cite{Xu:2009:DLS:1629575.1629587,Xu:2009:OSP:1674659.1677125,Fu:2009:EAD:1674659.1677044,mariani2008automated,jiang2008automatic}, testing~\cite{andrews1998_testing_using_log_file_analysis, elyasov2012_log_based_testing}, and failure diagnosis~\cite{syer2013leveraging,Nagaraj:2012:SCA:2228298.2228334}.

One of the challenges of leveraging the information in logs is to collect logs from end users of the software~\cite{hengtse}. In particular, developers of applications that run on end users' devices (\emph{e.g.} mobile apps) cannot directly retrieve the default logs. To address this challenge,  mobile analytics services enable developers to automatically record and transmit logging information using a centralised infrastructure. For example, Firebase provides APIs~\cite{firebase_log_events} that developers can use to send runtime information of an application to a cloud-based infrastructure for later analysis.

The unique nature of mobile analytics logging may lead to different logging practices compared to traditional logging that is stored locally. For example, mobile analytics logging may bring extra non-negligible resource overhead to generate and transmit the logs. In addition, the remote storage of information may also bring privacy concerns~\cite{10.1109/ASE.2019.00069}. 
Therefore, research into the characteristics of practices of mobile analytics logging is germane 
to help practitioners and researchers further understand and address the challenges. 

We conducted an empirical study on the use of the most widely adopted mobile analytics logging library, Firebase Analytics, in 57 open-source Android projects. We aim to answer two research questions:

\begin{description}[style=unboxed, leftmargin=0em]

\item[\textbf{RQ$_1$}] \textit{What are the characteristics of logging practices with mobile analytics?}

\item[\textbf{RQ$_2$}] \textit{What do developers log with mobile analytics?}

\end{description}

In our empirical study, we disclosed several interesting findings.
We found logging statements of mobile analytics are less pervasive and less maintained than traditional local logging statements. 
We also identified four types of information are usually logged: domain/business events, user interface events, failures and/or unexpected situations, and other information. Notably, the majority of log statements record domain/business events.
This paper presents our first investigation on the characteristics of mobile analytics logging practices and makes the following contributions:
\begin{itemize}[leftmargin=*, topsep=2pt, itemsep=2pt]
    \item Our paper is the first to study the logging practices with mobile analytics. 
    \item The data from our study, such as the custom logging classes we identified, provides a head-start to support further research on logging with mobile analytics. 
    \item We provide a comprehensive reproduction package\footnote{
    \url{https://github.com/mobileanalyticslogs/mobileanalyticslogging}}.
\end{itemize}

\section{Background: mobile analytics}
\label{sec:bg}
Mobile analytics combines software code that runs on mobile devices, data collection from the mobile device to servers, and processing and subsequent analysis. The data may include device data, app usage data, contextual information,~\emph{etc.}~\cite{spivey2010embedded_mobile_analytics_in_a_mobile_device_patent}. The analysis is performed by one or more stakeholders, such as the app's development team, the provider of the analytics, and/or the provider of the platform, ~\emph{e.g.} Android and iOS.

Mobile analytics was derived from web analytics and intended to focus on business-oriented metrics and reporting. Over time mobile analytics was also used for other purposes and their APIs evolved to support the collection of additional forms of information, including those mentioned earlier. One product in particular, known as Firebase Analytics, evolved and was acquired by several companies in turn as more and more developers chose to incorporate it into their mobile apps. Google now owns it, and Firebase Analytics are incorporated into over 62\% of the Android apps in the Google Play store~\cite{appbrain_firebase}. One of the key features from a developer-oriented point of view was the flexible and powerful APIs that enable developers to design and implement custom events and data collection, including log messages~\cite{firebase_log_events}. 

Mobile analytics has similarities to logging where developers write code statements that use a software library to record information while the software runs. However much of the logging developers use is intended for local consumption, for instance on a mobile device connected to their development machine,~\emph{i.e. local logging}. Remote logging copies what was written to a local log and sends it to a remote server for further use. There are various libraries available for mobile apps~\emph{e.g.}~Timber, however none is very popular~\cite{appbrain_logging_libraries}. Key differences between using logging and analytics include the integral analytical aspects of processing the data.

\section{Case study}
\label{sec:study}


In this section, we present our case study and the results through answering our two research questions.

\noindent \textbf{Subjects.} Our study is based on analyzing 25,611 Java projects used in prior research of logging utilities practices~\cite{boyuanlogutility}. In particular, all the Java projects are obtained from the GHTorrent~\cite{10.5555/2664446.2664449} MySQL dump  (last updated on 2019-06-01). Duplicate (\textit{e.g.} forks and clones) projects were removed as were inactive projects. In order to identify the projects that use Firebase Analytics, we further filtered the projects by searching keyword ``FirebaseAnalytics'' in each project's source code and collected 107 projects. For each of the 107 projects we manually examined whether customised APIs are used by developers for mobile analytics logging. We found 50 of them only collect the default automated metrics generated by Firebase without proactively logging any information. Therefore, our study focuses on the remaining 57 projects where developers intentionally leverage the logging features in Firebase 
to collect information for their software.

\noindent \textbf{Identifying logging statements.} Initially we followed the same practice as prior research where logging statements are identified based on the Abstract Syntax Tree (AST) and the particular pattern of method invocation of each logging library~\cite{yizengemse,chen2017characterizing}. 
However, after manually examining the logging statements, we found \emph{developers often wrap the Firebase APIs in a custom logging class}. In particular, logging using the Firebase API is rather complex where multiple method invocations are often needed to 
log once. Therefore, developers rarely directly call the Firebase APIs to log. Instead, they create wrapper class that provides utility methods that log using the Firebase API.
We manually identified these wrapper classes and their utility methods. These methods were used as keywords to automatically identify the logging statements in their respective project.
Specifically, we leveraged srcML to convert source code files to XML files. (Kotlin source files were first renamed from \texttt{*.kt} to \texttt{*.java} which was enough to process them). We then extracted all the invocation calls by using XPath. Then, for each project we used regular expressions to check whether the caller names matched with the corresponding class names we found.

\subsection*{RQ$_1$: What are the characteristics of logging practices with mobile analytics?}

\subsubsection*{Motivation}
Prior research studied characteristics of logging code in open-source server and desktop software~\cite{yuan2012characterizing, chen2017characterizing}, and mobile applications \cite{yizengemse}. However, these studies only focused on local logging code. We conjecture the unique characteristics of mobile analytics logging practices may introduce new challenges and opportunities for researchers and practitioners who design infrastructures for mobile analytics logging. In this RQ, we explore the characteristics of mobile analytics logging as well as how it evolves over time.

\subsubsection*{Approach}
For each of the 57 projects, we followed prior studies on logging practices~\cite{yuan2012characterizing, chen2017characterizing, yizengemse, shang2015studying, Kabinna2018emse} that study the logging practices based on the following metrics: 

    \noindent \textbf{SLOC per logging statement(Code density of logging).} The code density of logging, defined by Yuan~\emph{et al}~\cite{yuan2012characterizing}, is calculated by $SLOC/NOL$. SLOC refers to the number of source code lines and NOL refers to the number of logging lines. This metric measures the density of the mobile analytics logging statement in the projects. We used \emph{cloc} to count the number of source lines of code.

    \noindent \textbf{Churn rate of logging code.} This metric is calculated by averaging the $LogChurn/NOL$ of all the commits. LogChurn is calculated by the sum of added, deleted and modified logging statements.
    It measures the maintenance effort of the mobile analytics logging statements. For each of the projects, we analyzed its entire commit history and identified the added, deleted, and updated files in every commit by querying the GitHub API. For the added and deleted files we applied the same approach in previous step to detect the logging lines in those files. For the updated files we leveraged~\texttt{git diff} to identify the added logging lines, deleted logging lines, and updated logging lines and categorised them accordingly. 

\subsubsection*{Results}

\begin{figure*}
  \centering
    \includegraphics[width=0.85\linewidth]{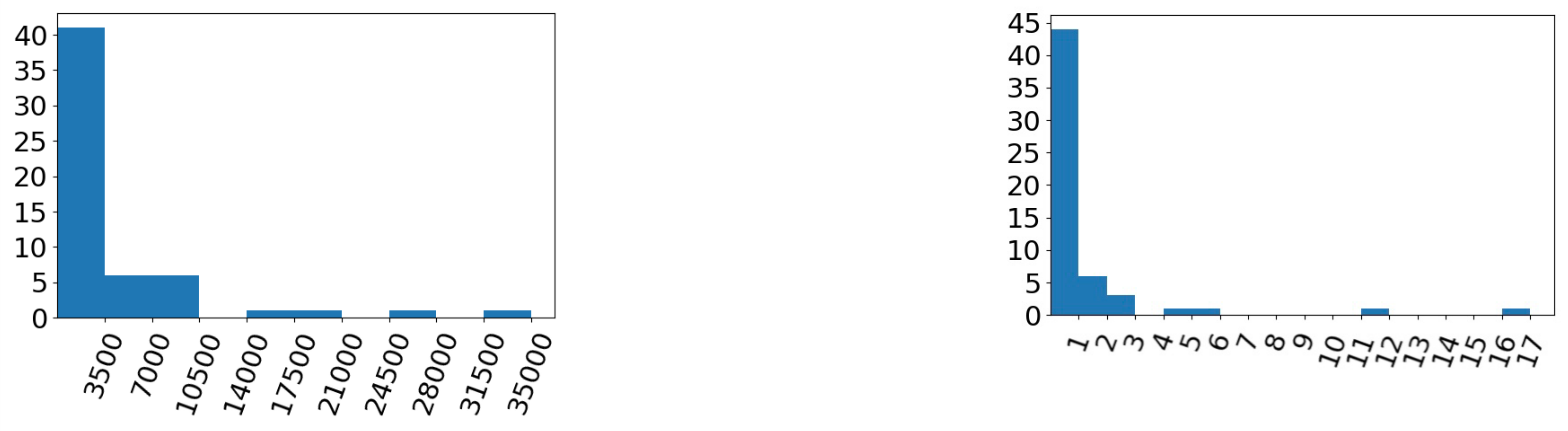}
  \caption{The characteristics of mobile analytics logging. (a) Code density of logging. (b) Logging code churn rate (\%).}
  \vspace{-0.5cm}
  \label{fig:metrics}
\end{figure*}

\textbf{Mobile analytics logging is less pervasive and less maintained than traditional mobile logging.} The median SLOC per mobile analytics logging statement is 1,148, while a prior study on FDroid apps showed the median value of SLOC was 145 per traditional mobile logging statement~\cite{yizengemse}. 

We have two hypotheses for the lower density of log statements using mobile analytics: \textit{1)} Developers may need to conservatively choose where and what to log in order to avoid impacting the end users' experiences from the mobile analytics logging (they incur both performance and bandwidth overhead because the logs have to be transmitted over a functional network connection). \textit{2)} The latency of deploying apps to end users and the impact on end users of debugging in the field reduces the temptation to use mobile analytics for debugging purposes.

In addition, among all the code commits, only 1.35\% (1,331 out of 98,565) contain changes (adding deleting or modifying) to mobile analytics logging statements, while a prior study shows around 10\% of the commits containing changes to traditional mobile logging statements~\cite{yizengemse}. By further examining the data, 
we find 56\% of the changes add, 23.6\% delete, and  20.3\% modify mobile analytics logging statements. 
In contrast, the prior study found a much larger amount of the changes to traditional mobile logging statements were to delete existing logging statements that were used for temporary debugging~\cite{yizengemse}. The different purposes of logging statements using mobile analytics may contribute to the differences of maintenance activities of mobile analytics logging statements (\emph{c.f.} RQ2). 

Figure~\ref{fig:metrics} presents histograms of mobile analytics logging statement density and churn rate of this code among the 57
apps. Although the majority of the apps in the study have both low log density and log churn rate, the long tail of both figures demonstrates some projects extensively use and maintain their mobile analytics logging statements.
For example,~\href{https://github.com/opendocument-app/OpenDocument.droid}{\texttt{OpenDocument}} has only around 4.2K SLOC yet it has 64 mobile analytics logging statements. By manually checking those projects we found 
developers sometimes log every UI interaction of the app in order to know how the app is being used. 
The high density may not lead to high churn rate of mobile analytics logging. By calculating the Pearson correlation between the two metrics, we found the two metrics are only weakly correlated ($\rho=-0.15$), 
which means the volume of mobile analytics logging statements may be independent with how often those logging statements are revised during ongoing development. 
For example,~\href{https://github.com/romannurik/muzei}{\texttt{muzei}} has 69 mobile analytics logging statements.
Only 20 of the 2,664 commits in the project's 7 year history mention the addition or revision of analytics. 









\subsection*{RQ$_2$: What do developers log in mobile analytics?}

\subsubsection*{Motivation}
Here, 
we study what types of information developers log using mobile analytics. Understanding precisely what is logged may help the mobile community design and build more useful 
mobile analytics frameworks (for instance relevant APIs) and infrastructures (for instance that optimises timely data delivery and analysis).

\subsubsection*{Approach}

We manually analysed a set of 300 randomly selected logging statements (CI=5, CL=95\%)~\cite{sample_size_calculator}. These are our data points here. Our method was as follows:


\emph{Step 1: Agreement on an initial code book}: We randomly selected 30 logging statements (\textit{i.e.} 10\% of the entire set) from the sample. All the researchers individually analysed the logging statements. The researchers then used the logging statements (and the source code of the project when needed) to assign a single label to each data point. These labels indicate what that statement logs. Once we individually labeled these 30 logging statements we compared our labels one-by-one with the goal of devising an initial shared code book.


\emph{Step 2: Analysis of the entire dataset}: We then randomly divided the 300 logging statements in five batches of 60 logging statements each. Each researcher was assigned to two batches. In other words, every logging statement is reviewed by two researchers. Using the initial code book and shared knowledge, the researchers individually labelled all the data points in the dataset.

\emph{Step 3: Resolving the disagreements}: After the first round, we observed agreements ranging between 66\% and 77\% among the researchers. Each pair of researchers would then discuss their disagreements, item by item, and reach an agreement. If the two researchers still did not agree a third researcher joined the discussion (which happened solely for two logging statements). During this resolution phase the researchers decided not to label five data points due to lack of clarity on the logging statement.


\subsubsection*{Results}

Four distinct types of information are being logged:



\noindent \textbf{Domain/business events (51.3\%)}: The majority of the mobile analytics logging statements are about specific domain or business events that just happened in the app. For example, the~\href{https://github.com/edx/edx-app-android}{\texttt{edX}} mobile app logs whether a user downloaded a video in their platform, and the~\href{https://github.com/blockchain/My-Wallet-V3-Android}{\texttt{MyWallet}} app tracks whether the user purchased something again.

\noindent \textbf{User interface events (38.7\%)}: Developers log the interactions users have with their app's interface, \textit{e.g.} when someone clicks at a button, selects an item in a menu, or opens a new activity. As examples, the~\href{https://github.com/blockchain/My-Wallet-V3-Android}{\texttt{AndroidBible}} app logs when users click the search button, and the~\href{https://github.com/StepicOrg/stepik-android}{\texttt{Stepik Android}} app logs when users visit the launch screen or select a course in the list of courses.

\noindent \textbf{Failures and/or unexpected situations (6.3\%)}: Interestingly, only a small portion of the mobile analytics logging statements focus on failures and unexpected situations, such as network errors, or a failure in a process. For example,~\href{https://github.com/opendocument-app/OpenDocument.droid}{\texttt{OpenDocument}} logs whenever it fails to open a file, and~\href{https://github.com/blockchain/My-Wallet-V3-Android}{\texttt{AndroidBible}} logs when a network error occurs.

\noindent \textbf{Others (2\%)}: Finally, we observed a few mobile analytics logging statements that focused on the app version or aspects of the users. For example, the~\href{https://github.com/OneBusAway/onebusaway-android}{\texttt{OneBusAway}} app logs the region of the user.

Our results show \textbf{developers choose to log different information with mobile analytics than with traditional logging libraries}. A prior study on logging practices in Android apps found half of local logging statements were used for debugging purposes~\cite{yizengemse}. Similar findings are also reported by recent studies on server application logs~\cite{hengtse,chen2017characterizing}. However, the information logged using mobile analytics rarely assists in debugging, as most of the information is collected to understand the behaviour of end users. For example, the information about domain and business events may help developers understand the usage of the features of their application; and information about the user interface events helps understand the habits of users' interactions with the applications. On the other hand, intuitively, the information that helps in debugging often contains lower-level details. Due to the potential performance and data bandwidth overhead of logging with mobile analytics, developers may choose not to record lower-level information even though it might be useful for debugging. Our results highlight the need for further infrastructure support for field debugging activities through logging in mobile analytics. 

\section{Threats to Validity}
\label{sec:threats}


\emph{Internal and construct validity.} We used static analysis to detect the logging lines (\emph{i.e.} calls to the mobile analytics framework). For each project, we manually identified the keywords that were used to detect mobile analytics calls, 
however, we may have missed log statements 
that have since been renamed or removed.
Moreover, given the static nature of our tool we may have missed some method calls. Nonetheless, the manual analysis we conducted after the data collection did not reveal any 
omissions. 
Therefore, we believe our choice of data collection tool did not adversely influence the results. 

The different types of information observed in the mobile logs were manually extracted from 300 randomly selected logging statements. To reduce possible biases, each logging line was coded and agreed by two researchers, and the final code book was agreed by all the authors of this paper. We also make our raw data, analysis, and code book available online.


\emph{External validity.} We studied 57 open-source mobile applications. Larger studies, including both iOS and proprietary mobile software, are needed before we can claim our results are generalizable to the entire population of mobile apps. Nevertheless, we argue our results provide solid initial results on how developers use mobile analytics in their Android apps.

\section{Future Work}
\label{sec:future}

Our 
study can be extended in at least two directions.

\textit{Characterizing contextual information recorded in the logs.}
We categorised the logs based on the types of the logged events.
Developers can also log contextual information of the events.
For example, when the \texttt{OpenDocument} app logs failures to open a file it also records the file type.
Such contextual information is presumably of interest to the developers.
We propose a qualitative study would help characterise the logged contextual information and also examine the relationships between different types of events and their corresponding contextual information.

\textit{Understanding developers' perceptions on using mobile analytics.}
We analyzed the source code of Android apps to characterise the use of mobile analytics.
However, it is difficult to know the developers' actual intentions to use mobile analytics versus local logging merely by analyzing the code.
Interviews with Android developers may provide a better understanding of when and why developers would choose to log using mobile analytics.

\section{Related Work}
\label{sec:related}

Empirical studies have been conducted on the general practices of logging. Yuan \emph{et al.}~\cite{yuan2012characterizing} conducted the first quantitative empirical study on logging practices, which focused on C and C++ projects. To complement this pioneering study, Chen \emph{et al.}~\cite{chen2017characterizing} and Zeng \emph{et al.}~\cite{yizengemse} studied the logging practices in Java projects and Android app projects, respectively. To further understand the decisions of logging in practice,  a recent qualitative study by Li~\emph{et al.}~\cite{hengtse} investigated the benefits and costs of logging by interviewing developers and studying logging-related issue reports. Besides those characteristic studies, research often studied particular aspects in logging, including their evolution and stability~\cite{Kabinna2018emse, shang2013jsep}, data leakage~\cite{zhou2020_mobilogleak},  
the logging libraries~\cite{kabinna2016logging}, their utilities~\cite{boyuanlogutility}, and logging configurations~\cite{Zhi2019AnES}. Finally, the relationship between software quality, performance, error-handling, and logging practices are also important aspects that are studied in prior research~\cite{shang2015studying,Chowdhury2017AnES,yizengemse,Oliveira_Borges_Silva_Cacho_Castor_2018_android_error_handling}. 
Prior studies, e.g., study by Zeng \emph{et al.}~\cite{yizengemse}, do not consider logging practices where the information is recorded and transmitted to a centralised, remote, system using mobile analytics. 



Mobile analytics tools are widely deployed to monitor software at runtime in the field. They have also been used to identify ways to improve app quality~\cite{harty2020improving} and software testing~\cite{harty2015mobile}. Some of the mobile analytics tools focus on monitoring performance of the software~\cite{HASSELBRING2020100019,10.1145/2901739.2901774}. 
A recent study found mobile analytics are often used poorly and with user-privacy concerns~\cite{10.1109/ASE.2019.00069}. 

\section{Conclusion}
\label{sec:conclusion}
In this paper, we conducted an empirical study on the use of the most popular mobile analytics framework, \textit{i.e.} Firebase, to perform logging in 57 open-source Android projects. We observed distinct practices of logging practices with mobile analytics. In particular, logs in mobile analytics are less pervasive and less maintained than traditional, local, logging practices. The most common information being logged using mobile analytics are domain/business events; while almost none of the mobile analytics logs focus on debugging, unlike traditional, local, logs.

Given the popularity of using mobile analytics in both open-source and closed-source apps there is much to gain and learn from investigating how mobile analytics are used in those apps.
We plan to further extend the study by a) further studying the contextual information logged for the different types of events, b) understanding developers' intentions to decide when to use mobile analytics, and c) combining additional qualitative studies together with developer interviews to better characterise the logging practices with mobile analytics.

\section*{Acknowledgments}
NII Shonan provided a stimulating environment for collaboration~\cite{nii_shonan_152}. Lili Wei was supported by the Postdoctoral Fellowship Scheme of the Hong Kong Research Grant Council.


\bibliographystyle{IEEEtran}
\bibliography{IEEEabrv,main.bib}

\end{document}